\documentclass[pre,superscriptaddress,floatfix,twocolumn,showpacs]{revtex4-2}
\usepackage{verbatim}
\usepackage{amsmath}
\usepackage{amsfonts}
\usepackage{amssymb}
\usepackage{graphicx}
\usepackage{bm}
\usepackage{color}
\usepackage{hyperref}
\usepackage[active]{srcltx}
\usepackage{cases}
\usepackage{ulem}
\usepackage{multirow}
\usepackage{braket}
\usepackage{colortbl}

\def\stacksymbols #1#2#3#4{\def\theguybelow{#2}
    \def\verticalposition{\lower#3pt}
    \def\spacingwithinsymbol{\baselineskip0pt\lineskip#4pt}
    \mathrel{\mathpalette\intermediary#1}}
\def\intermediary#1#2{\verticalposition\vbox{\spacingwithinsymbol
      \everycr={}\tabskip0pt
      \halign{$\mathsurround0pt#1\hfil##\hfil$\crcr#2\crcr}}}

\begin{document}
\title{Fate of Topological Edge States in Disordered Periodically-driven Nonlinear Systems}

\author{Ken Mochizuki}
\affiliation{Advanced Institute for Materials Research (WPI-AIMR), Tohoku University, Sendai 980-8577, Japan}

\author{Kaoru Mizuta}
\affiliation{Department of Physics, Kyoto University, Kyoto 606-8502, Japan}

\author{Norio Kawakami}
\affiliation{Department of Physics, Kyoto University, Kyoto 606-8502, Japan}


\begin{abstract}
We explore topological edge states in periodically driven nonlinear systems. Based on a self-consistency method adjusted to periodically driven systems, we obtain stationary states associated with topological phases unique to Floquet systems. In addition, we study the linear stability of these edge states and reveal that Floquet stationary edge states experience a sort of transition between two regions I and II, in which lifetimes of these edge states are extremely long and short, respectively. We characterize the transitions in lifetimes by Krein signatures or equivalently the pseudo-Hermiticity breaking, and clarify that the transitions between regions I and II are signified by collisions of edge-dominant eigenstates of Floquet operators for fluctuations. We also analyze the effects of random potentials and clarify that lifetimes of various stationary edge states are equalized due to the randomness-induced mixing of edge- and bulk-dominant eigenstates. This intriguing phenomenon originating from a competition between the nonlinearity and randomness results in that random potentials prolong lifetimes in the region II and vice versa in the region I. These changes of lifetimes induced by nonlinear and/or random effects should be detectable in experiments by preparing initial states akin to the edge states. 
\end{abstract}


\maketitle
\section{introduction}
\label{sec:introduction}
Periodically driven systems or equivalently Floquet systems have attracted much attention. This is because topological phases of matter can be manipulated by utilizing periodic driving, which is referred to as the Floquet engineering \cite{oka2009photovoltaic,
kitagawa2011transport,
oka2019floquet,
harper2020topology,
rudner2020band}.
A variety of phenomena originating from Floquet topological phases have been observed in numerous settings, such as electronic systems \cite{wang2013observation,mciver2020light}, cold atoms  \cite{jotzu2014experimental}, and photonic systems \cite{rechtsman2013photonic,
mukherjee2017experimental,
maczewsky2017observation,
chen2018observation}. 
While the Floquet engineering is based on analogies to topological phases of static systems in many cases, it has also been revealed that topological phases which have no static counterpart can exist owing to the periodicity in time \cite{kitagawa2010topological,
asboth2013bulk,
rudner2013anomalous,
asboth2014chiral,
asboth2015edge,
nathan2015topological,
titum2016anomalous,
morimoto2017floquet,
roy2017floquet,
higashikawa2019floquet,
mochizuki2020bulk,
mochizuki2020topological}. Such topological phases have been recognized as unique phases in nonequilibrium systems, often referred to as anomalous Floquet topological phases.

In addition to the periodic driving mentioned above, various physical effects which can make systems nonequilibrium, such as gain-loss and nonlinearity, can be present in many physical settings. Even under such effects, topological edge states and related phenomena have been observed \cite{poli2015selective,
xiao2017observation,
weimann2017topologically,
mukherjee2017experimental,
chen2018observation,
harari2018topological,
bandres2018topological,
maczewsky2020nonlinearity,
mukherjee2020observation,
mukherjee2020observation_arXiv}. 
Especially, signatures of edge states in the nonlinear regime are quite intriguing, since topological numbers are defined in linear systems and thus relations between nonlinear phenomena and topological phases are nontrivial. Therefore, topological aspects of various nonlinear phenomena have been extensively explored  
\cite{lumer2013self,
ablowitz2014linear,
ablowitz2015strong,
gerasimenko2016attractor,
hadad2016self,
leykam2016edge,
solnyshkov2017chirality,
malzard2018nonlinear,
pal2018amplitude,
bisianov2019stability,
chaunsali2019self,
kruk2019nonlinear,
sone2019anomalous,
wang2019topologically,
ivanov2020edge,
maczewsky2020nonlinearity,
mukherjee2020observation,
mukherjee2020observation_arXiv,
mochizuki2020stability,
smirnova2020nonlinear,
sone2020exceptional,
ablowitzs2021peierls,
chaunsali2021stability}.

In this paper, we focus on periodically driven nonlinear systems described by the Gross-Pitaevskii equation and obtain Floquet stationary states which originate from topological edge states characterized by the Rudner winding number \cite{rudner2013anomalous}. While solitons and their relation to topological phases have been studied in similar settings \cite{lumer2013self,
leykam2016edge,
maczewsky2020nonlinearity,
mukherjee2020observation,
mukherjee2020observation_arXiv}, we obtain stationary edge states which are directly related to anomalous Floquet topological phases in a manner different from solitons. We also analyze the stability of these edge states and find that they undergo a sort of transition between the long-lifetime region (I) and the short lifetime regions (II) with varying the strength of nonlinearity, which is experimentally detectable.  We clarify that the transitions are signified by collisions of eigenvalues with opposite Krein signatures or equivalently the pseudo-Hermiticity breaking of edge-dominant eigenstates for fluctuations. This suggests that nonlinear systems will provide an intriguing playground to study the pseudo-Hermiticity breaking other than open systems in which essentially the same or similar transitions have been extensively studied \cite{bender1998real,
bender2002generalized,
mostafazadeh2002pseudoI,
mostafazadeh2002pseudoII,
mostafazadeh2002pseudoIII,
mostafazadeh2003exact,
mostafazadeh2004pseudounitary,
guo2009observation,
ruter2010observation,
chtchelkatchev2012stimulation,
regensburger2012parity,
peng2014loss,
peng2014parity,
feng2014single,
hodaei2014parity,
poli2015selective,
ashida2017parity,
xiao2017observation,
el2018non,
longhi2018parity,
kawabata2019symmetry}. We also reveal that lifetimes of the edge states exhibit universal behavior when random potentials exist since the edge- and bulk-dominant eigenstates are mixed, leading to that lifetimes are prolonged by random potentials in the region II and shortened in the region I.  While the robustness of topological edge states against randomness is often discussed, it is an intriguing phenomenon that random potentials tend to stabilize edge states in the region II.

The rest of this paper is organized as follows. In Sec. \ref{sec:model_formalism}, we introduce a model which describes periodically driven nonlinear systems, summarize unique features of Floquet stationary states, and explain how to analyze their stability. In Sec. \ref{sec:g-dependence}, we numerically obtain Floquet stationary states and show that these states are inherited from topological edge states in the linear regime. Through the linear stability analysis of the edge states, we uncover a sort of transition in lifetimes and characterize them by Krein signatures or the pseudo-Hermiticity breaking of edge-dominant eigenstates. In Sec. \ref{sec:v-dependence}, we explore the effects of random potentials and reveal the equalization of lifetimes for various edge states as a result of the competition between nonlinear and random effects. Section \ref{sec:summary} is devoted to summary.

\begin{figure*}[t]
\begin{center}
\includegraphics[scale=1.7]{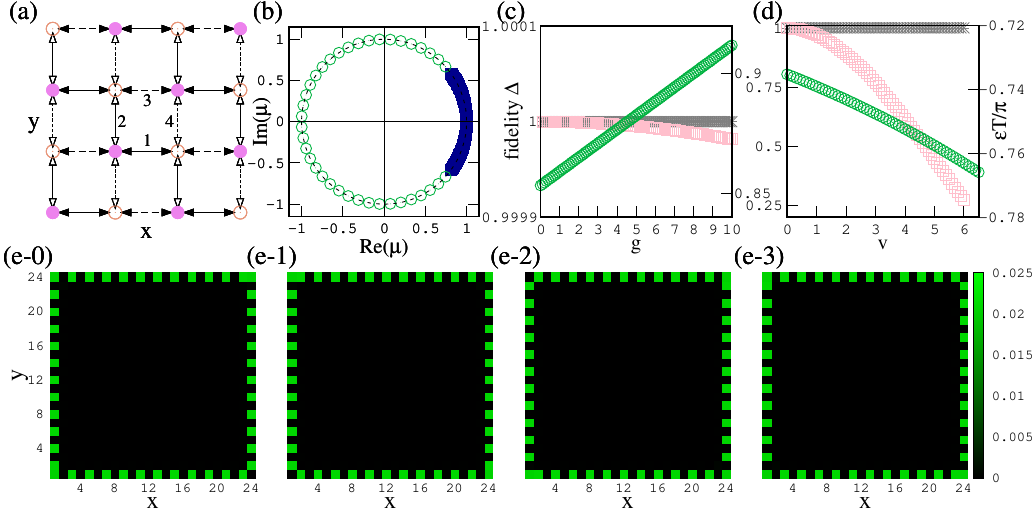}
\caption{(a) A schematic picture of the hopping terms $H_m$ in four time frames $m$ in Eq. (\ref{eq:Hamiltonian}). Solid filled, solid empty, broken filled, and broken empty arrows are respectively couplings for $m=1,2,3,4$. (b) Eigenvalues of $\mathbb{T}\exp[-i\int_0^TdtH(t)\,]$ in the linear regime ($g=0$) under OBC, with $h=2.2\pi/T,\,T=1,\,L=24$, and $V=0$. Green empty circles correspond to topological edge states inside the gap of the bulk spectrum composed of navy filled squares. A few topological edge states in the bulk spectrum are included in navy filled squares. (c) Fidelities between Floquet stationary edge states (left axis) and quasienergies (green circles, right axis) as functions of $g$ with $V(\bm r)=0$. Gray asterisks and pink empty squares respectively represent $\Delta[\phi_{g-\delta g}(0),\phi_g(0)]$ and $\Delta[\phi_{g=0}(0),\phi_g(0)]$. We choose one stationary state $\ket{\phi_g(t)}$ that is continuously connected to the topological edge state with $\varepsilon T/\pi\simeq0.85$ in the linear regime $g=0$. (d) Fidelities between Floquet stationary edge states $\ket{\phi_v(t)}$ (left axis) and quasienergies (green circles, right axis) as functions of $v$ at $g=-3$. Gray asterisks and pink empty squares show $\Delta[\phi_{v-\delta v}(0),\phi_v(0)]$ and $\Delta[\phi_{v=0 }(0),\phi_v(0)]$ respectively. (e) Intensity profiles of an edge state $|\phi({\bm r},t)|^2$ corresponding to the state in (c), with $g=4,\,h=2.2\pi/T,\,T=1$, and $V=0$ at (e-0) $t=0$ (or $T$), (e-1) $t=T/4$, (e-2) $t=T/2$, and (e-3) $t=3T/4$.}
\label{fig:schematics_eigenvalue_fidelity_intensity}
\end{center}
\end{figure*}

\section{model and formulation}
\label{sec:model_formalism}
In this section, we explain a model which describes the dynamics of periodically driven nonlinear systems and how to analyze stationary states and their stability in such systems.
\subsection{Nonlinear 2D systems under periodic driving}
\label{subsec:model}
We consider two-dimensional systems on a square lattice $\Lambda=\{ (x,y) \, | \, x,y=1,2,\hdots,L \}$, in which a state is written as
\begin{align}
\ket{\psi(t)}=\sum_{\bm r}\psi({\bm r},t)\ket{\bm r},
\label{eq:state}
\end{align}
with the position ${\bm r} \in \Lambda$ and the time $t$. 
We focus on systems governed by the Gross-Pitaevskii equation
\begin{align}
i\frac{\partial \psi({\bm r},t)}{\partial t}=
\sum_{{\bm r}'}H_{{\bm r},{\bm r}'}(t)\psi({\bm r}',t)
+g|\psi({\bm r},t)|^2\psi({\bm r},t),
\label{eq:model}
\end{align}
where $H_{{\bm r},{\bm r}'}(t+T)=H_{{\bm r},{\bm r}'}(t)$ is satisfied. Experiments in which the dynamics of light is described by Eq. (\ref{eq:model}) have been carried out for arrayed optical fibres modulated periodically along the direction of light propagation  \cite{ablowitz2014linear,maczewsky2020nonlinearity,mukherjee2020observation,mukherjee2020observation_arXiv}. In such cases, each fiber corresponds to each lattice position ${\bm r}=(x,y)$. Kerr effects in optical fibers can generate the nonlinear term $g|\psi({\bm r},t)|^2\psi({\bm r},t)$. During the dynamics, the total intensity $P=\sum_{\bm r}|\psi({\bm r},t)|^2$ is conserved. Throughout this paper, we fix $T=1$ and $P=1$ without loss of generality and impose open boundary conditions (OBC). The former corresponds to introducing dimensionless quantities: $t,\,\psi({\bm r},t),\,H_{{\bm r},{\bm r}'}(t),\,g \rightarrow t/T,\,\psi({\bm r},t)/\sqrt{P},\,H_{{\bm r},{\bm r}'}(t)T,\,gTP^2$. Regarding the linear term in the right hand side of Eq. (\ref{eq:model}), we consider the modulation which is separated into four time frames,
\begin{align}
H(t)=H_m+V\ \ \text{for}\ \ (m-1)T/4 \le t < mT/4,
\label{eq:Hamiltonian}
\end{align}
where $H(t)=\sum_{{\bm r},{\bm r}'}H_{{\bm r},{\bm r}'}(t)\ket{\bm r}\bra{\bm r'}$ and $m=1,2,3,4$. The modulation terms $H_m$ describe evanescent couplings between the nearest lattices, which are depicted in Fig. \ref{fig:schematics_eigenvalue_fidelity_intensity} (a); $H_1,\,H_2,\,H_3$, and $H_4$ include hopping terms corresponding to solid filled, solid empty, broken filled, broken empty arrows, respectively. In a time frame $(m-1)T/4 \le t < mT/4$, hopping terms other than $H_m$ are zero. For simplicity, we take the same value for hoppings in all time frames and write it as $h$,
\begin{align}
H_m=h\sum_{\bm r}\ket{\bm r}\bra{{\bm r}_m}
\label{eq:hopping}
\end{align}
where ${\bm r}_m$ represents the site linked to ${\bm r}$ in $(m-1)T/4 \le t < mT/4$, as explained above. We also apply on-site random potentials given by 
\begin{align}
V=\sum_{\bm r}V(\bm r)\ket{\bm r}\bra{\bm r}.
\label{eq:potential}
\end{align}
In the case of optical fibers, such position dependent potentials correspond to refractive indices which are different in each fiber under the paraxial approximation. When we consider the situation in which $h \simeq 2\pi/T$, $g=0$, and $V=0$, the Floquet system described by $H(t)$ accommodates topological edge states characterized by the Rudner winding number $W=1$  \cite{rudner2013anomalous}. The nontrivial Rudner winding number, related to the winding of quasienergies $i\log(\mu)/T$ where $\mu$ represents eigenvalues of the Floquet operator in the linear regime, results from the periodicity in time and thus the corresponding edge states are unique to Floquet systems. Green empty circles in Fig. \ref{fig:schematics_eigenvalue_fidelity_intensity} (b) represent a part of such topological edge states which reside in the gap of bulk states.

\subsection{Scheme for the linear stability analysis of Floquet stationary states}
\label{subsec:stationary-states_stability}
In the periodically driven nonlinear systems introduced above, we explore Floquet stationary states which satisfy
\begin{align}
\ket{\phi(T)}=F[\{\phi(t)\}]\ket{\phi(0)}
=e^{-i\varepsilon T}\ket{\phi(0)},
\label{eq:stationary_state}
\end{align}
where $F[\{\psi(t)\}]$ is defined as
\begin{align}
F[\{\psi(t)\}]&=\mathbb{T}\exp[-i\int_0^TdtH_g(t)\,],
\label{eq:time-evolution_operator}\\
H_g(t)&=H(t)+\sum_{\bm r}g|\psi({\bm r},t)|^2
\ket{\bm r}\bra{{\bm r}},
\label{eq:H_tilde}
\end{align}
with $\mathbb{T}$ being the time ordering operator. A Floquet stationary state earns the phase $e^{-i \varepsilon T}$ during the one nonlinear cycle $T$, where $\varepsilon$ is referred to as the quasienergy. 

From the stationary states in Eq. (\ref{eq:stationary_state}), we can construct Floquet states $\ket{\phi_\text{F}(t)}$ with the periodicity $T$,
\begin{align}
\ket{\phi(t)}=e^{-i \varepsilon t}\ket{\phi_\text{F}(t)},\ \ 
\ket{\phi_\text{F}(t+T)}=\ket{\phi_\text{F}(t)}.
\label{eq:floquet_state}
\end{align}
As is shown below, the dynamics of $\ket{\phi_\text{F}(t)}$ during one period determines the stability of the corresponding Floquet stationary state. We separate the state at time $t$ into a stationary state and fluctuations from it,
\begin{align}
\ket{\psi(t)}=e^{-i \varepsilon t}
[\,\ket{\phi_\text{F}(t)}+\ket{\delta\psi(t)}\,],
\label{eq:separation}
\end{align}
where $|\delta\psi({\bm r},t)|/|\phi_\text{F}({\bm r},t)|\ll1$ is supposed. Substituting Eq. (\ref{eq:separation}) into Eq. (\ref{eq:model}) and ignoring secondary and higher order terms, we obtain the linear time-evolution equation of $\ket{\delta\psi(t)}$
\begin{align}
i \frac{\partial}{\partial t} \ket{\delta \Psi(t)} &= K(t) \ket{\delta\Psi(t)}, \quad K(t)=K(t+T),
\label{eq:dynamics_delta-psi}\\
\ket{\delta\Psi(T)} &=G[\{\phi_\text{F}(t)\}]\ket{\delta\Psi(0)},
\label{eq:one-cycle_delta-psi}
\end{align}
where $\ket{\delta\Psi(t)}$ includes $\ket{\delta\psi(t)}$ and its complex conjugate,
\begin{align}
\ket{\delta\Psi(t)}=\left[\begin{array}{c}
\ket{\delta\psi(t)}\\
\ket{\delta\psi(t)}^\ast
\end{array}\right].
\label{eq:linearized_psi}
\end{align}
The non-Hermitian operator $K(t)$ and the nonunitary operator $G[\{\phi_\text{F}(t)\}]$ are defined as
\begin{align}
K(t)&=\left[\begin{array}{cc}
Q(t)&R(t)\\
-R^\ast(t)&-Q(t)
\end{array}\right],
\label{eq:K}\\
Q(t)&=H(t)+\sum_{\bm r}\ket{\bm r}\bra{\bm r}
[2g|\phi_\text{F}({\bm r},t)|^2-\varepsilon],
\label{eq:Q}\\
R(t)&=\sum_{\bm r}\ket{\bm r}\bra{\bm r}g\phi_\text{F}^2({\bm r},t),
\label{eq:R}\\
G[\{\phi_\text{F}(t)\}]&=\mathbb{T}\exp[-i\int_0^TdtK(t)\,].
\label{eq:time-evolution_operator_linearized}
\end{align}
The Floquet state $\phi_\text{F}({\bm r},t)$ returns to itself after one period $T$ and plays the role of a time-periodic potential in  $K(t)$, thereby making $K(t)$ a time-periodic non-Hermitian operator. Equation (\ref{eq:dynamics_delta-psi}) says that the dynamics of $\delta\Psi({\bm r},t)$ becomes equivalent to that of a linear open Floquet system under the non-Hermitian ``Hamiltonian" $K(t)$, with the doubled system size of the original lattice $\Lambda$. This results in the non-unitary Floquet operator  $G[\{\phi_\text{F}(t)\}]$, defined for the dynamics of $\delta\Psi({\bm r},t)$, and hence the norm of $\ket{\delta\Psi(t)}$ is not conserved in general. When $\ket{\delta\Psi(t)}$ grows with time, then $\ket{\psi(t)}$ in Eq. (\ref{eq:separation}) largely deviates from $\ket{\phi(t)}$ and the stationary state is unstable. In such a situation, if we prepare an initial state akin to $\ket{\phi_\text{F}(0)}$, this state has a finite lifetime and collapses as time evolves. Whether $\ket{\delta\Psi(t)}$ grows or not is determined from the eigenvalues $\lambda_n$ of $G[\{\phi_\text{F}(t)\}]$,
\begin{align}
G[\{\phi_\text{F}(t)\}]
\ket{\delta\Phi_\text{F}^n}
=\lambda_n\ket{\delta\Phi_\text{F}^n},
\label{eq:eigen_equation}
\end{align}
where $\ket{\delta\Phi_\text{F}^n}$ is the corresponding eigenstate which describes the dynamics of fluctuations. When all $\lambda_n$ satisfy $|\lambda_n|<1$, $\ket{\delta\psi(t)}$ decays with $t$ and then the Floquet stationary state is stable; $\ket{\psi(t)}$ converges to $\ket{\phi(t)}$ as time evolves. If there exists $\lambda_n$ whose absolute value is larger than $1$, the perturbation $\ket{\delta\psi(t)}$ grows and $\ket{\phi(t)}$ is unstable. In the present case, if there is an eigenvalue $\lambda_n$, then $\lambda_n^\ast$ and $(\lambda_n^\ast)^{-1}$ always exist. This is because $G[\{\phi_\text{F}(t)\}]$ satisfies 
\begin{align}
\sigma_3G^{-1}[\{\phi_\text{F}(t)\}]\sigma_3
&=G^\dagger[\{\phi_\text{F}(t)\}],
\label{eq:pseudo-Hermiticity}\\
\sigma_1G^\ast[\{\phi_\text{F}(t)\}]\sigma_1
&=G[\{\phi_\text{F}(t)\}],
\label{eq:particle-hole_symmetry}
\end{align}
resulting from $\sigma_3K(t)\sigma_3=K^\dagger(t)$ and $\sigma_1K^\ast(t)\sigma_1=-K(t)$  where $\sigma_1$ and $\sigma_3$ are Pauli matrices in the space spanned by $\ket{\delta\psi(t)}$ and $\ket{\delta\psi(t)}^\ast$. Therefore, eigenvalues of $G[\{\phi_\text{F}(t)\}]$ either come in pairs of $|\lambda_n|>1$ and $|\lambda_n|<1$ or satisfy $|\lambda_n|=1$. The state $\ket{\phi(t)}$ is unstable in the former case. In the latter case, the linear stability analysis cannot clarify whether $\ket{\psi(t)}$ approaches $\ket{\phi(t)}$ or not, although there is not such a situation and stationary states are unstable in the present work. In the presence of symmetries in Eqs. (\ref{eq:pseudo-Hermiticity}) and (\ref{eq:particle-hole_symmetry}), the eigenvalues of $G[\{\phi_\text{F}(t)\}]$ can be characterized by Krein signature \cite{arnold1968ergodic,
flynn2020deconstructing,
zhang2020pt}
\begin{align}
S_n=\text{sign}(\bra{\delta\Phi_\text{F}^n}
\sigma_3\ket{\delta\Phi_\text{F}^n}).
\label{eq:krein-signature}
\end{align}
The Krein signature $S_n$ becomes $\pm1$ or $0$ where the former and latter cases respectively correspond to $|\lambda_n|=1$ and $|\lambda_n|\neq1$. As we will see momentarily below, with a small change of parameters in $G[\{\phi_\text{F}(t)\}]$, two eigenvalues on the unit circle with opposite Krein signatures $S_n=\pm1$ need to collide for the emergence of eigenvalues outside the unit circle with $S_n=0$ \cite{arnold1968ergodic,
flynn2020deconstructing,
zhang2020pt}.

\section{nonlinear effects: an emergent transition in lifetimes}
\label{sec:g-dependence}
We first consider cases in which the on-site random potential is absent $V=0$. We numerically obtain Floquet stationary states which are associated with anomalous Floquet topological phases in the linear regime. In particular, we clarify that such stationary states show transitions in lifetimes with respect to the strength of nonlinearity $g$.

\subsection{Floquet stationary edge states}
\label{subsec:derivation}
We begin with the numerical method for obtaining Floquet stationary states inherited from topological edge states. Here, we combine a self-consistency method for Floquet stationary states in Ref. \cite{lumer2013self} with the fidelity so that we can associate them to topological edge states. First, we choose an initial state $\ket{\psi(0)}$ and obtain $\ket{\psi(t)}$ by simulating the nonlinear dynamics of one cycle. Second, we construct $F[\{\psi(t)\}]$ in Eq. (\ref{eq:time-evolution_operator}) and diagonalize it. Among the eigenstates of $F[\{\psi(t)\}]$, which we write as $\ket{\psi_\text{F}}$, we take a state closest to $\ket{\psi(0)}$. We measure the similarity between a pair of states $\ket{\psi}$ and $\ket{\phi}$ based on the fidelity
\begin{align}
\Delta(\psi,\phi)=\frac{|\langle\psi|\phi\rangle|}{\sqrt{\langle\psi|\psi\rangle\langle\phi|\phi\rangle}}.
\label{eq:fidelity}
\end{align}
 Third, choosing $\ket{\psi_\text{F}}$ as the initial state, which has the largest $\Delta[\psi(0),\psi_\text{F}]$, we repeat the first and second processes. Until the fidelity approaches $1$, we iterate these procedures. The linear system at $g=0$ is topologically nontrivial, thereby having edge states $\ket{\phi_\text{edge}(t)}$ characterized by the nontrivial Rudner winding number $W=1$ under OBC \cite{rudner2013anomalous}. If we choose one of topological edge states $\ket{\phi_\text{edge}(t=0)}$ as $\ket{\psi(0)}$ for the first iteration, it is expected that a stationary state similar to the topological edge state is obtained for a small $g$. Actually, we can obtain $\ket{\phi_{\delta g}(t)}$ which satisfies $\Delta[\phi_{\delta g}(0),\phi_\text{edge}(0)]\simeq1$ with $\delta g=\pm0.1$, where $\ket{\phi_g(t)}$ is the Floquet stationary state when the strength of nonlinearity is $g$. In this way, we gradually alter $g$ from $0$ and numerically obtain stationary edge states in the nonlinear regime, taking the edge state at $g-\delta g$ as the initial state for the first iteration at each $g$. Then, the obtained states follow the change of $g$. 

Numerical results are shown in Fig. \ref{fig:schematics_eigenvalue_fidelity_intensity} (c) and (e0)-(e3). Note that we focus on topological edge states whose eigenvalues are inside the gap of bulk states, which are described by green empty circles in Fig. \ref{fig:schematics_eigenvalue_fidelity_intensity} (b), since it is comparatively difficult to obtain Floquet stationary states originating from other edge states whose eigenvalues are close to those of bulk states. For several parameter sets adopted, starting from topological edge states inside the gap, $\Delta[\phi_{g-\delta g}(0),\phi_g(0)]$ is almost always larger than $0.9999$ in the gradual change of $g$ by the step width $\delta g=\pm0.1$ within $|g|\leq10$. Figure \ref{fig:schematics_eigenvalue_fidelity_intensity} (c) shows $\Delta[\phi_{g-\delta g}(0),\phi_g(0)]$, $\Delta[\phi_{g=0}(0),\phi_g(0)]$, and the quasienergy $\varepsilon$ at each $g$ when we obtain stationary states based on the procedure explained above starting from one of topological edge states. The fidelities close to $1$ and the continuity of quasienergies establish that the obtained stationary states are directly connected to topological edge states in the linear regime $g=0$. Figure \ref{fig:schematics_eigenvalue_fidelity_intensity} (e) shows an edge state obtained from a topological edge state with $\varepsilon T/\pi\simeq0.85$. The intensity profile becomes the same as the initial one after one period $T$, while it changes during the dynamics. Note that, in contrast to solitons in similar systems explored in previous studies \cite{lumer2013self,
leykam2016edge,
maczewsky2020nonlinearity,
mukherjee2020observation,
mukherjee2020observation_arXiv}, the states which we have obtained are more directly related to topological edge states; The stationary states are reduced to topological edge states in the limit of $g\rightarrow0$, while solitons are not.

\begin{figure}[thbp]
\begin{center}
\includegraphics[scale=0.9]{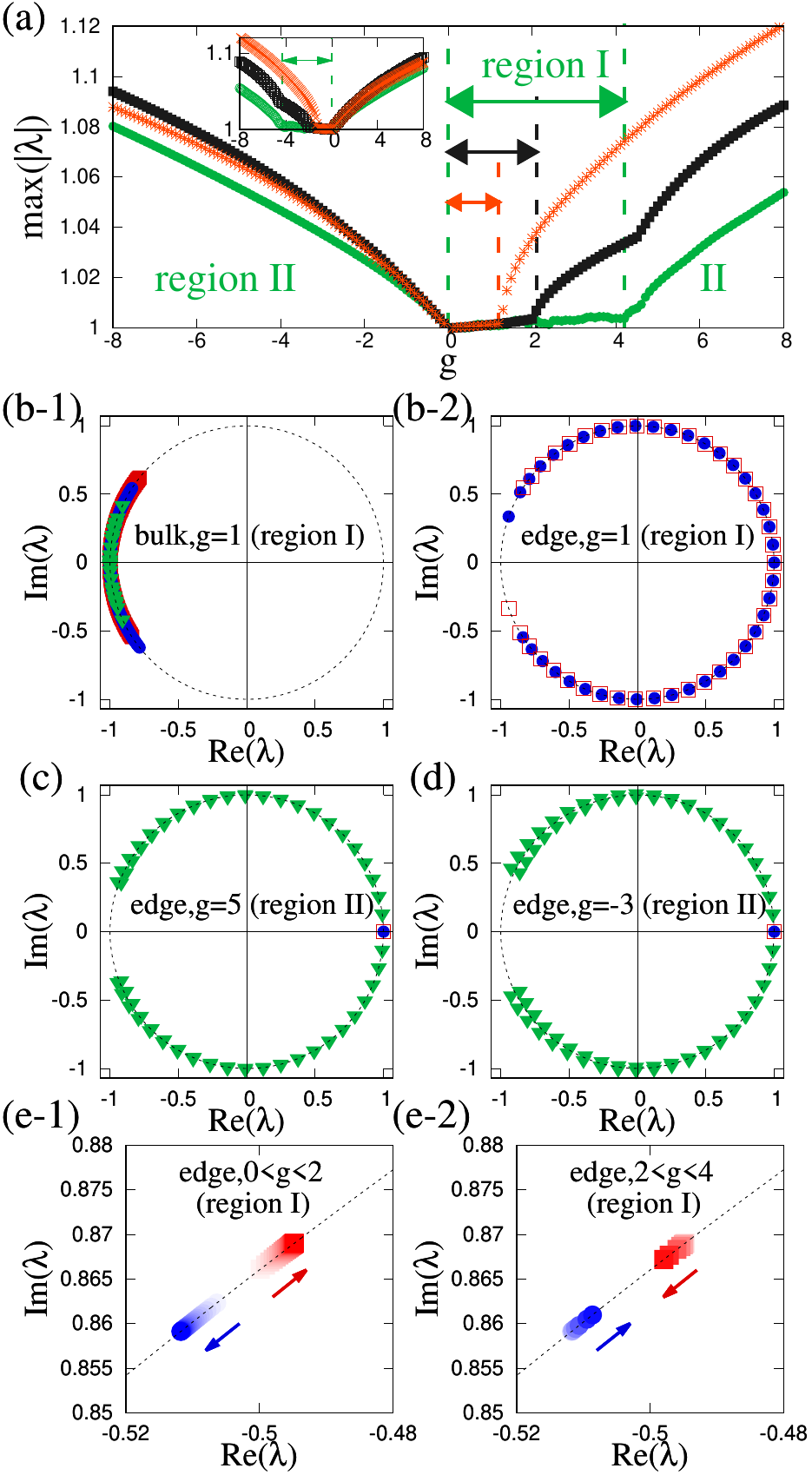}
\caption{(a) Dependence of $\max (|\lambda|)$ on the strength of nonlinearity $g$ with $h=2.2\pi/T$ and $L=24$ for Floquet stationary states originating from topological edge states whose quasienergies at $g=0$ are $\varepsilon T/\pi\simeq0.98$\,(green filled circles), $0.85$\,(black filled squares), and $0.43$\, (orange asterisks). Each stationary state has two distinct regions, the long-lifetime region I and the short-lifetime region II. The inset figure shows the results for states with negative quasienergies at $g=0$; $\varepsilon T/\pi\simeq-0.98$\,(green empty circles), $-0.85$\,(black empty squares), and $-0.43$ (orange crossess). (b)-(e) Eigenvalues of $G[\{\phi_\text{F}(t)\}]$ with various $g$ for $\ket{\phi_\text{F}(t)}$ that has the largest region I, corresponding to green circles in (a). Green triangles, red squares, and blue circles represent eigenvalues whose Krein signatures are $0$, $+1$, and $-1$, respectively. The dashed curves show the unit circle. (b) Eigenvalues at $g=1$, where $P_\text{edge}^n<0.5$ in (b-1) and $P_\text{edge}^n\geq0.5$ in (b-2). (c),(d) Eigenvalues with $g=5$ and $g=-3$ respectively , where the corresponding eigenstates satisfy $P_\text{edge}^n\geq0.5$. (e) Trajectories of two eigenvalues with $P_\text{edge}^n\geq0.5$ in (e-1) $0<g<2$ and (e-2) $2<g<4$. Symbols with light (deep) color correspond to smaller (larger) $g$ and eigenvalues move in the directions of arrows as $g$ is increased.}
\label{fig:max_eigenvalue}
\end{center}
\end{figure}

\subsection{Transitions in lifetimes}
\label{subsec:stability}
 Let us now explore the stability of the obtained Floquet stationary edge states based on the formalism given in Sec. \ref{subsec:stationary-states_stability}. Figure \ref{fig:max_eigenvalue} (a) shows max$(|\lambda|)$ of $G[\{\phi_\text{F}(t)\}]$ for Floquet states $\ket{\phi_\text{F}(t)}$ related to several topological edge states with different quasienergies, from which we find that the behavior of max$(|\lambda|)$ depends on the signs of $g$ and $\varepsilon$. In Fig. \ref{fig:max_eigenvalue} (a) we can see sharp transitions in lifetimes with increasing $g$; max$(|\lambda|)$ for an edge state little increases with increasing $g$ from $0$ but sharply increases if $g$ is over a threshold value. As a result, there is a region in which max$(|\lambda|)$ is close to $1$ and thus the corresponding edge states have longer lifetimes, which we refer to as the region I. In the region II, outside the region I, lifetimes of edge states are short in comparison to that in region I. Thus, the linear stability analysis concludes transition-like behaviors in the stability of stationary states as far as they are inherited from topological edge states. The width of the region I changes according to quasienergies, where the edge state whose original quasienergy at $g=0$ is closest to $\pi$ has the largest width.  We note that there is no transition in lifetimes if $\varepsilon$ at $g=0$ is positive (negative) and $g$ is negative (positive), as shown in Fig. \ref{fig:max_eigenvalue} (a) and its inset. A similar behavior is also observed in a different geometry where periodic and open boundary conditions are imposed respectively in $x$ and $y$ directions, and thus we expect that transitions of lifetimes occur in various geometries not restricted to the present situation. 
 
 The transitions can be characterized by Krein signatures in Eq. (\ref{eq:krein-signature}) and edge weights of eigenstates defined below. As mentioned in Sec. \ref{subsec:stationary-states_stability}, the emergence of pair eigenvalues $|\lambda_n|>1$ and $|\lambda_n|<1$ is triggered by a collision of two eigenvalues with opposite Krein signatures. This is because $G[\{\phi_\text{F}(t)\}]$ satisfies Eqs. (\ref{eq:pseudo-Hermiticity}) and (\ref{eq:particle-hole_symmetry}). It is also important for the stability whether or not eigenstates have large amplitudes at the edge, since the Floquet stationary states are confined to the edge of the system. For evaluating to what extent eigenstates are confined to the edge, we introduce edge weights
 \begin{align}
 P_\text{edge}^n=\frac{
 \bra{\delta\Phi_\text{F}^n}
 \hat{P}_\text{edge}
\ket{\delta\Phi_\text{F}^n}} {\braket{\delta\Phi_\text{F}^n|
 \delta\Phi_\text{F}^n}},\,\,
 \hat{P}_\text{edge}
 =\sum_{\tilde{\bm r}}
 \ket{\tilde{\bm r}}
 \bra{\tilde{\bm r}},
 \label{eq:P_edge}
 \end{align}
where the sum for $\tilde{\bm r}$ is taken along the edge; $\tilde{\bm r}=(x=1,1 \leq y \leq L),\,(x=L,1 \leq y \leq L),\,(1 \leq x \leq L,y=1)$, and $(1 \leq x \leq L,y=L)$. Figure \ref{fig:max_eigenvalue} (b-1) and (b-2) show eigenvalues of $G[\{\phi_\text{F}(t)\}]$ respectively with $P_\text{edge}^n<0.5$ and $P_\text{edge}^n\geq0.5$ for the stationary edge state which resides in the region I of long lifetimes. We can clearly see from Fig. \ref{fig:max_eigenvalue} (b) that only eigenvalues of bulk-dominant eigenstates ($P_\text{edge}^n<0.5$) deviate from the unit circle or equivalently their Krein signatures are zero in the long-lifetime region. On the other hand, in the region II, Krein signatures of edge-dominant eigenstates ($P_\text{edge}^n\geq0.5$) become zero, which we can understand from Fig. \ref{fig:max_eigenvalue} (c) and (d). Therefore, it is concluded that the transitions between regions I and II are signified by collisions of eigenvalues with large $P_\text{edge}^n$. We can indeed confirm the above statement by observing the motions of eigenvalues. Figure \ref{fig:max_eigenvalue} (e) shows trajectories of two eigenvalues with $P_\text{edge}^n\geq0.5$ and opposite Krein signatures when $g$ is increased within the region I. For small $g$, the two eigenvalues with $S_n=\pm1$ move in the opposite directions and go away from each other. After the repulsion, the directions of motions are reversed, the eigenvalues get close to each other, and finally they undergo a collision, resulting in shorter lifetimes in the region II. 

The collisions of eigenvalues with opposite signatures elucidated above are directly related to the breaking of the pseudo-Hermiticity. This is because Eq. (\ref{eq:pseudo-Hermiticity}) corresponds to the pseudo-Hermiticity $\sigma_3H_\text{F}\sigma_3=H_\text{F}^\dagger$ of the corresponding Floquet ``Hamiltonian'' $H_\text{F}=i\log\left(G[\{\phi_\text{F}(t)\}]\right)/T$ \cite{mostafazadeh2002pseudoI,
mostafazadeh2002pseudoII,
mostafazadeh2002pseudoIII,
mostafazadeh2003exact,
mostafazadeh2004pseudounitary,
esaki2011edge,
lieu2018topological,
kawabata2019symmetry,
zhang2020pt}. Note that, while the condition $\eta H_\text{F} \eta^{-1}=H_\text{F}^\dagger$ with a positive definite operator $\eta$ is equivalent to the reality of all eigenvalues of $H_\text{F}$ ($|\lambda_n|=1$ for all $n$) \cite{mostafazadeh2002pseudoI,
mostafazadeh2002pseudoII,
mostafazadeh2002pseudoIII,
mostafazadeh2004pseudounitary}, $\sigma_3$ is not positive and then $\sigma_3H_\text{F}\sigma_3=H_\text{F}^\dagger$ does not solely ensure $|\lambda_n|=1$. In the presence of the pseudo-Hermiticity in Eq. (\ref{eq:pseudo-Hermiticity}), the condition $\sigma_3\ket{\delta\Phi_\text{F}^n}\propto\ket{\delta\Upsilon_\text{F}^n}$ corresponds to $|\lambda_n|=1$ where $\bra{\delta\Upsilon_\text{F}^n}$ is the left eigenstate of $G[\{\phi_\text{F}(t)\}]$ whose eigenvalue is $\lambda_n$. When $|\lambda_n|\neq1$,  $\sigma_3\ket{\delta\Phi_\text{F}^n}$ is not proportional to the corresponding left eigenstate, which we refer to as the pseudo-Hermiticity breaking. Such transitions accompanied by collisions of eigenvalues have been extensively studied in open systems with gain and/or loss which are not restricted to pseudo-Hermitian systems but also include $\mathcal{PT}$ symmetric systems described by non-Hermitian operators \cite{bender1998real,
bender2002generalized,
mostafazadeh2002pseudoI,
mostafazadeh2002pseudoII,
mostafazadeh2002pseudoIII,
mostafazadeh2003exact,
mostafazadeh2004pseudounitary,
guo2009observation,
ruter2010observation,
chtchelkatchev2012stimulation,
regensburger2012parity,
peng2014loss,
peng2014parity,
feng2014single,
hodaei2014parity,
poli2015selective,
ashida2017parity,
xiao2017observation,
el2018non,
longhi2018parity,
kawabata2019symmetry}. In the present case where there is no gain/loss and the total intensity $P$ is conserved, the mathematical structure essentially the same or quite similar to open systems emerges through the linear stability analysis; Stationary states in the nonlinear dynamics induce non-Hermitian terms in the linearized time-evolution equations of fluctuations, as shown in Sec. \ref{subsec:stationary-states_stability}. Thus, the present nonlinear systems can be alternative platforms to explore intriguing transitions signified by eigenvalue collisions and symmetry breaking of eigenstates extensively discussed in open systems.

\begin{figure}[bt]
\begin{center}
\includegraphics[scale=0.85]{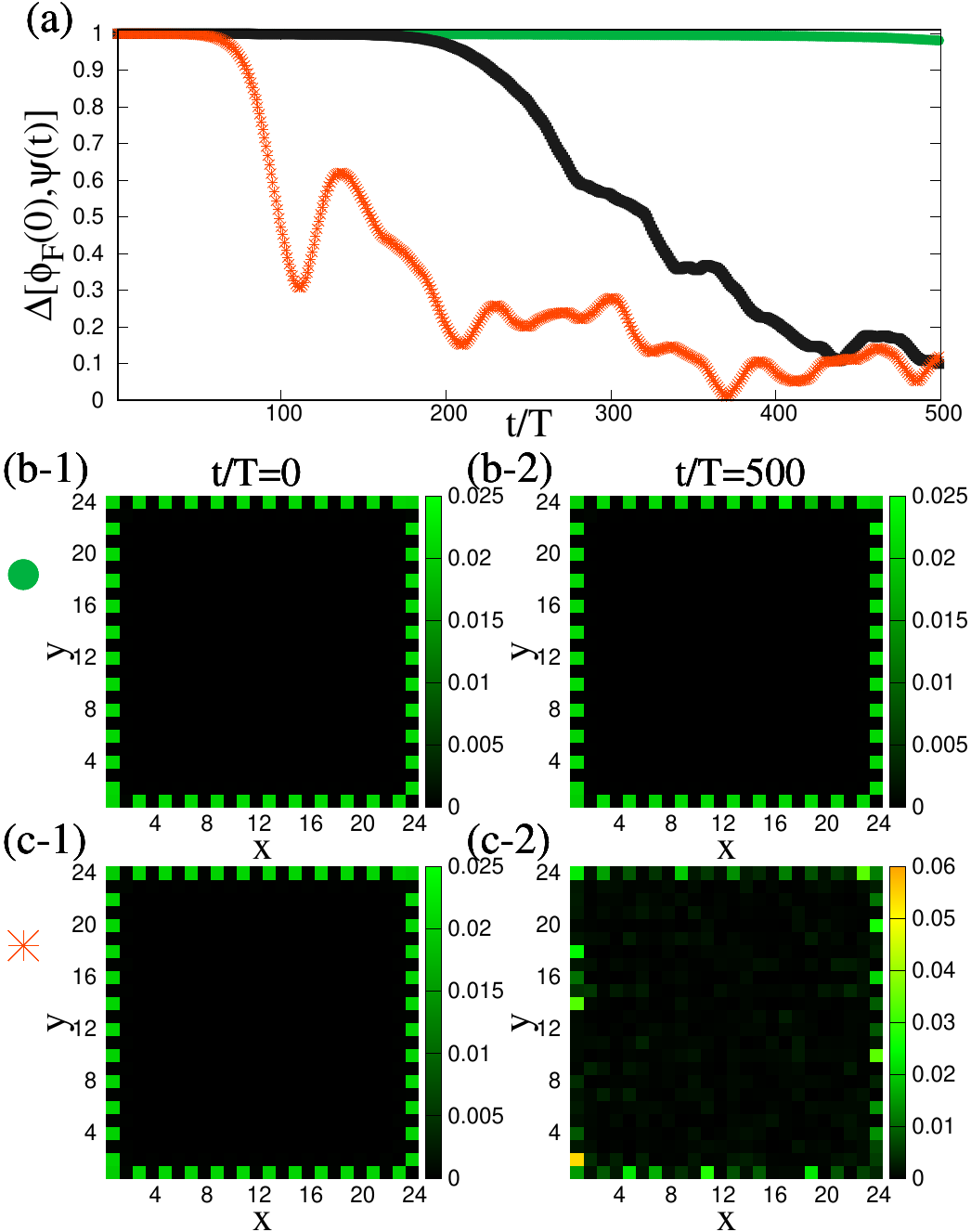}
\caption{(a) The fidelities between Floquet states $\ket{\phi_\text{F}(0)}$ and $\ket{\psi(t=mT)}$ with $m$ being integers when initial states are chosen as in Eq. (\ref{eq:initial_state}) with $g=4$ and $w=10^{-3}$. The meanings of symbols are the same as in Fig. \ref{fig:max_eigenvalue} (a). (b) The intensity profiles at (b-1) $t=0$ and (b-2) $t=500T$ corresponding to the dynamics shown by green filled circles in (a). (c) The same figure as (b) except that the dynamics corresponds to orange asterisks in (a). }
\label{fig:fidelity_intensity}
\end{center}
\end{figure}

 We can detect the above sharp transitions through the real-time dynamics of the fidelity $\Delta[\phi_\text{F}(0),\psi(t=mT)]$ ($m \in \mathbb{N}$), defined by Eq. (\ref{eq:fidelity}). When we choose an initial state with small fluctuations around the Floquet stationary state
\begin{align}
\psi({\bm r},0)\propto[\phi_\text{F}({\bm r},0)+\rho(\bm r)],\ \ 
\rho(\bm r)\in[-w/2,w/2],
\label{eq:initial_state}
\end{align}
where $\rho(\bm r)$ obeys the box distribution, $\Delta[\phi_\text{F}(0),\psi(t)]$ decreases as $t$ increases since $\ket{\phi_\text{F}(t)}$ is unstable. Note that, the initial state is properly normalized such that $\sum_{\bm r}|\psi({\bm r},0)|^2=1$ is satisfied. Figure \ref{fig:fidelity_intensity} (a) shows the fidelities as functions of $t$ with $g=4$ and $w=10^{-3}$, for several stationary edge states. In this case, a stationary state originating from a topological edge state near $\pi$ (green filled circles) resides in the region I with long lifetimes, while other states (black filled squares and orange asterisks) are in the region II with short lifetimes, which causes a huge difference in lifetimes for these states. In Fig. \ref{fig:fidelity_intensity} (a), the fidelity $\Delta[\phi_\text{F}(0),\psi(t)]$ for the edge state in the region I is kept close to $1$ while $\Delta[\phi_\text{F}(0),\psi(t)]$ for other states in the region II largely deviates from $1$, clearly featuring the transitions in lifetimes. This phenomenon should be detectable in photonic experiments with coherent light. The intensities at $t=0$ and $t=500T$ are respectively shown in Fig. \ref{fig:fidelity_intensity} (b-1),(c-1) and (b-2),(c-2), which are accessible quantities in photonic experiments. Regarding the edge state in the long-lifetime region corresponding to (b), the intensities are confined to the edge after long time-evolution, while the intensities exhibit a complicated profile in the case of the edge state corresponding to (c) in the short-lifetime region.

\begin{figure}[bthp]
\begin{center}
\includegraphics[scale=1.5]{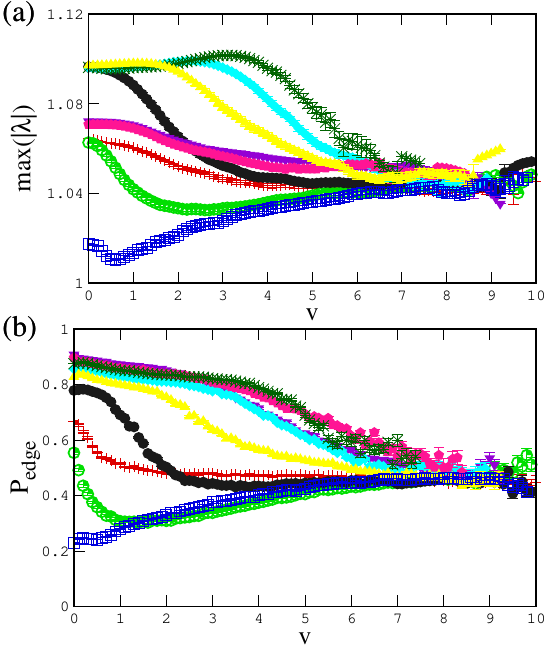}
\caption{(a) The largest values of $|\lambda_n|$ and (b) mean edge weights for eigenstates with $|\lambda_n|\neq1$ as functions of $v$, averaged over $100$ ensembles with $h=2.2\pi/T,\,g=-3$, and $L=12$. Dark green, light blue, yellow, black, purple, pink, red, light green, and blue symbols respectively correspond to various Floquet stationary edge states whose origins are different topological edge states with $\varepsilon T/\pi\simeq-0.53,\,-0.61,\,-0.70,\,-0.78,\,0.70,\,0.61,\,0.96,\,-0.87$, and $-0.96$. Under the nonlinearlity $g=-3$ in the clean limit $v=0$, the blue one represents the stationary edge state in the region I, while the other states reside in the region II.}
\label{fig:max_edge-ratio}
\end{center}
\end{figure}

\section{competition of the nonlinearity and randomness}
\label{sec:v-dependence}
\begin{figure}[bthp]
\begin{center}
\includegraphics[scale=1.5]{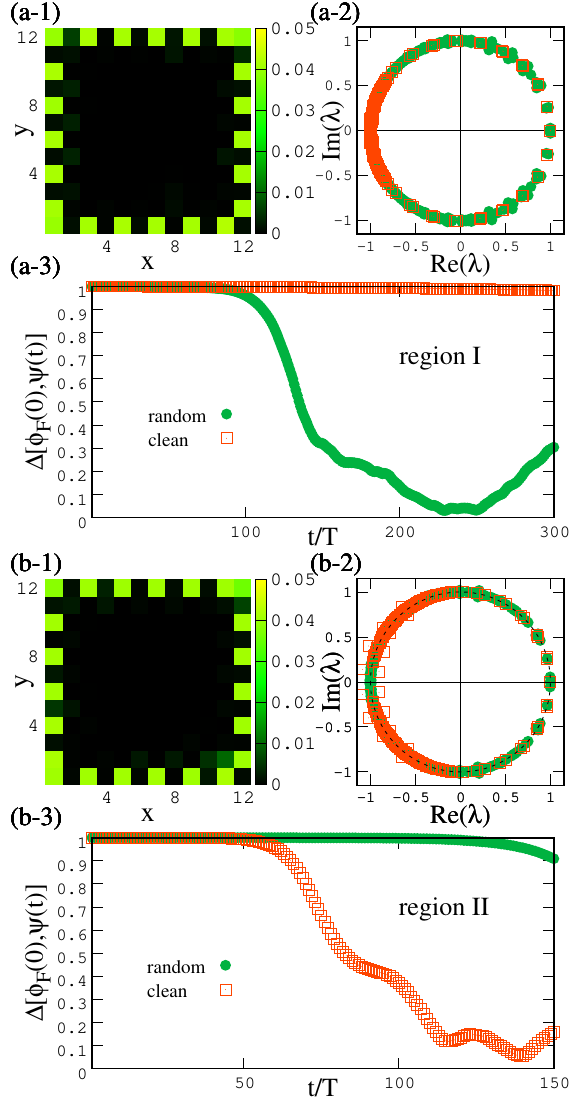}
\caption{(a-1,b-1) The intensities of Floquet stationary edge states at $t=0$, (a-2,b-2) corresponding eigenvalues of $G[\{\phi_\text{F}(t)\}]$, and (a-3,b-3) fidelities during the dynamics for initial states in Eq. (\ref{eq:initial_state}) with $w=10^{-3}$, under specific realizations of $V(\bm r)$. Stationary states in (a) and (b) respectively correspond to blue squares in the region I and yellow triangles in the region II in Fig. \ref{fig:max_edge-ratio}. In (a), green filled circles and orange empty squares correspond to $\ket{\phi_\text{F}(t)}$ respectively in random ($v=6.7$) and clean ($v=0$) systems, where (a-1) shows the intensity profiles in the former case. In (b), the meanings of symbols are the same as in (a) except that the strength of $V(\bm r)$ is $v=6.1$ in the random system. }
\label{fig:intensity_eigenvalue_fidelity}
\end{center}
\end{figure}

In this section, we explore effects of randomness and reveal the equalization of lifetimes where each Floquet stationary edge state looses its specific character due to the competition between the nonlinearity and randomness. We consider time-independent random potentials
\begin{align}
V(\bm r)\in[-v/2,+v/2]
\label{eq:random_potential}
\end{align}
in Eqs. (\ref{eq:Hamiltonian}) and (\ref{eq:potential}), where $v$ is the width of the box distribution. Even when $H(t)$ includes the random on-site potential $V$, the numerical method adopted in the previous section can be utilized. We obtain stationary edge states in a manner explained below. First, we fix a configuration of random numbers $\chi(\bm r)$ distributed between $-1/2$ and $1/2$, of which each number is allocated to each site. Second, we adopt a configuration of random potentials $V(\bm r)=\delta v\chi(\bm r)$ with a small magnitude $\delta v$, and obtain an edge state under the weak random potential. For the derivation, we take the  edge state with $g\neq0$ but without randomness $v=0$ as the initial state $\ket{\psi(0)}$ of the first iteration. Then, we can obtain $\ket{\phi_{v=\delta v}(t)}$ which satisfies $\Delta[\phi_{\delta v}(0),\phi_0(0)]\simeq1$, where $\ket{\phi_{v}(t)}$ is the edge state under the random potential $V(\bm r)=v\chi(\bm r)$. In the same way, we obtain $\ket{\phi_{v}(t)}$ utilizing $\ket{\phi_{v-\delta v}(0)}$ as the initial state for the iteration. When we slightly change the value of $v$ with $\delta v=0.1$, we adopt edge states satisfying $\Delta[\phi_{v}(0),\phi_{v-\delta v}(0)]>0.99$ and abolish other states not fulfilling the criterion, which ensures the connection between stationary edge states when $v$ is changed. Fidelities $\Delta[\phi_{v}(0),\phi_{v-\delta v}(0)]$, $\Delta[\phi_{v=0}(0),\phi_v(0)]$ and quasienergies $\varepsilon$ with various $v$ for obtaining $\ket{\phi_v(t)}$ in Fig. \ref{fig:intensity_eigenvalue_fidelity} (b-1) under a specific realization of $\chi(\bm r)$ are shown in Fig. \ref{fig:schematics_eigenvalue_fidelity_intensity} (d). While $\Delta[\phi_{v=0}(0),\phi_v(0)]$ deviates from $1$ for large $v$, fidelities and quasienergies continuously change as functions of $v$, which confirms the connection between these edge states.

We now discuss the stability of the obtained Floquet stationary edge states in random systems. Remarkably, max$(|\lambda|)$ for large random potentials are equalized for a variety of edge states. Figure \ref{fig:max_edge-ratio} (a) shows max$(|\lambda|)$ averaged over $100$ ensembles at each $v$ for several edge states originating from different topological edge states. As $v$ is increased, each edge state looses its specific character, and growth rates max$(|\lambda|)$ of fluctuations around various edge states converge to a constant value. Thus, if max$(|\lambda|)$ of an edge state  at $v=0$ is above (below) the convergent value, the growth rate decreases (increases) as $v$ is increased. In Fig. \ref{fig:max_edge-ratio}, results on one type of edge states whose lifetimes are shortened by random potentials are represented by blue empty squares. This state with $\varepsilon T/\pi\simeq-1$ at $g=0$ resides in the region I with largest width in terms of $g$, as clarified in Sec. \ref{subsec:stability}. On the other hand, lifetimes of all other states which reside in the region II are prolonged by randomness, and eventually approach approximately the same value as that of the stationary state in the region I after the strength of $V({\bm r})$ is sufficiently increased. 

The convergent behaviors of the lifetime observed both in the regions I and II can be attributed to the mixing of bulk-dominant eigenstates with small $P_\text{edge}^n$ and edge-dominant eigenstates with large $P_\text{edge}^n$ induced by random potentials. In the region I, weak instabilities are caused by bulk-dominant eigenstates at small $v$, but random potentials yield $\ket{\delta\Phi_\text{F}^n}$ with $|\lambda_n|\neq1$ and relatively large $P_\text{edge}^n$ through the mixing, which enhances max($|\lambda|$) at large $v$. On the other hand, in the region II, edge-dominant eigenstates mainly contribute to the instability in the clean system but the mixing by $V(\bm r)$ suppresses $P_\text{edge}^n$ for these states and then lifetimes are prolonged under strong random potentials. Actually, Fig. \ref{fig:max_edge-ratio} (a) and (b) show that the behavior of max($|\lambda|$) has an intimate relationship to that of $P_\text{edge}$. Edge weights $P_\text{edge}$ averaged over $\ket{\delta\Phi_\text{F}^n}$ with $|\lambda_n|\neq1$ and $100$ realizations of $\chi(\bm r)$ converge to a constant value for large $v$, giving rise to the convergence of max($|\lambda|$). 

The characteristic behavior of max$(|\lambda|)$ is also associated with the competition between the nonlinearity and randomness. In the present model, if we substitute $\phi_{v=0}({\bm r},t)$ into $G[\{\phi_\text{F}(t)\}]$ and increase the strength of randomness $v$, max$(|\lambda|)$ approaches $1$ since the random potential $V(\bm r)$ dominates the non-Hermitian term of $K(t)$ in Eq. (\ref{eq:R}) originating from the nonlinear effects $g|\psi({\bm r},t)|^2\psi({\bm r},t)$. This naturally explains the supression of the instability in the region II. Note that a similar behavior is commonly observed in a related but different context; In linear systems described by non-Hermitian operators, strong random potentials overcome non-Hermitian terms and suppress the growth rate of eigenstates \cite{hatano1996localization, hatano1997vortex,hatano1998non}. 
In our case, however, we have to further notice that what actually occurs is that $\phi_\text{F}({\bm r},t)$ is altered with increasing $v$. Although how $\phi_\text{F}({\bm r},t)$ is deformed by random potentials is not obvious, the fluctuation of $\phi_\text{F}({\bm r},t)$ would possibly lead to larger values of max$(|\lambda|)$ through the fluctuation of the non-Hermitian term $R(t)$, which could be a reason for the enhanced instability in the region I.  As a result of such a competition between the randomness and nonlinearity, max($|\lambda|$) converges to a constant value for large $v$.

 The randomness-induced enhancement and suppression of instabilities in regions I and II can be confirmed through the dynamics of the fidelity defined in Eq. (\ref{eq:fidelity}) for the initial state in Eq. (\ref{eq:initial_state}). Intensity profiles of Floquet stationary edge states in random systems are shown in Fig. \ref{fig:intensity_eigenvalue_fidelity} (a-1) and (b-1), respectively corresponding to blue squares and yellow triangles in Fig. \ref{fig:max_edge-ratio}. From Fig. \ref{fig:intensity_eigenvalue_fidelity} (a-1) and (b-1), we can see that the stationary states in random systems are well confined to the edge of the system. Orange empty squares and green filled circles in Fig. \ref{fig:intensity_eigenvalue_fidelity} (a-2) show eigenvalues of $G[\{\phi_\text{F}(t)\}]$ respectively with $v=0$ and $v=6.7$, where $\ket{\phi_\text{F}(t)}$ resides in the region I. From Fig. \ref{fig:intensity_eigenvalue_fidelity} (a-2), we can find that the instability at $v=0$ is extremely weak and the largest value of $|\lambda_n|$ under random potentials is larger than that without randomness. Orange empty squares and green filled circles in Fig. \ref{fig:intensity_eigenvalue_fidelity} (a-3) show fidelities at $t=mT$ in clean and random systems respectively, where $w$ is $10^{-3}$ and $m$ takes integer values. Comparing fidelities with and without random potentials, we can understand that the lifetime of the edge state is shortened by random potentials in the region I. On the other hand, in the region II, Fig. \ref{fig:intensity_eigenvalue_fidelity} (b-2) clearly shows that max($|\lambda_n|$) with randomness at $v=6.1$ is smaller than that in the clean system at $v=0$. Thus, $\ket{\psi(t=mT)}$ in the random system is kept close to $\ket{\phi_\text{F}(t=0)}$ after the fidelity $\Delta[\phi_\text{F}(0),\psi(t=mT)]$ in the clean system largely deviates from $1$ as shown in Fig. \ref{fig:intensity_eigenvalue_fidelity} (b-3), implying that lifetimes of edge states are prolonged by random potentials in the region II.

\section{summary}
\label{sec:summary}
We have studied Floquet stationary states in periodically driven nonlinear systems which have anomalous Floquet topological phases in the linear regime. By gradually altering the strength of nonlinearity, we have numerically obtained stationary states which are directly connected to anomalous topological edge states. 

Regarding the edge states, we have carried out the linear stability analysis and found that these edge states experience a sort of transition in lifetimes;  Stationary edge states have the parameter regions I and II with extremely long and short lifetimes, respectively. The transitions in lifetimes should be experimentally detectable by measuring the intensities of classical light. We have characterized the transitions of lifetimes in terms of Krein signatures or equivalently the pseudo-Hermiticity breaking, and clarified that collisions of edge-dominant eigenstates cause the transitions between regions I and II. The eigenvalue collisions accompanied by the symmetry breaking of eigenstates have a quite similar mathematical structure to those studied in open systems in the linear regime \cite{bender1998real,
bender2002generalized,
mostafazadeh2002pseudoI,
mostafazadeh2002pseudoII,
mostafazadeh2002pseudoIII,
mostafazadeh2003exact,
mostafazadeh2004pseudounitary,
guo2009observation,
ruter2010observation,
chtchelkatchev2012stimulation,
regensburger2012parity,
peng2014loss,
peng2014parity,
feng2014single,
hodaei2014parity,
poli2015selective,
ashida2017parity,
xiao2017observation,
el2018non,
longhi2018parity,
kawabata2019symmetry,
zhang2020pt}, although the periodically driven nonlinear systems studied here are not open systems since the total intensity $P$ is conserved during the dynamics. In the present case, the structure appears in the linear stability analysis, suggesting that nonlinear systems can give platforms different from open linear systems for the observation of the pseudo-Hermiticity breaking and the related transitions. 

In addition, we have explored the effects of randomness and revealed that the instabilities are suppressed (enhanced) by random potentials in the region II (I). This is because random potentials cause the mixing of edge- and bulk-dominant eigenstates, leading to the same convergent value of the growth rate for fluctuations, no matter which edge state is chosen. Since edge-dominant eigenstates drastically shorten lifetimes, and instabilities are mainly caused by edge (bulk) eigenstates in the region II (I) with no randomness, the mixing of these eigenstates leads to the suppression (enhancement) of instabilities in the region II (I). While the robustness of topological edge states against random potentials is often discussed, randomness creates the tendency to stabilize edge states in the region II, which is an intriguing phenomenon due to the competition between the nonlinear and random effects.

\section*{acknowledgement}
This work was supported by JSPS KAKENHI Grants No. JP18J20727, JP19H01838, JP20H01845, and JP20J12930. Kaoru Mizuta also appreciates the support of WISE Program from MEXT.
\bibliographystyle{apsrev4-2}
\bibliography{reference.bib}

\end{document}